\documentclass[prb,twocolumn,showpacs,floats,amsmath,amssymb]{revtex4}
\usepackage[dvips]{graphicx}

\def\bra#1{{\langle #1 \vert}}
\def\ket#1{{\vert #1 \rangle}}
\def\x#1{{\sigma_{#1}^{x}}}
\def\z#1{{\sigma_{#1}^{z}}}
\def\y#1{{\sigma_{#1}^{y}}}

\begin{document}

\title{Quantum renormalization group of XYZ model
in a transverse magnetic field}

\author{A. Langari}
\affiliation{
Institute for Advanced Studies in Basic Sciences, Zanjan 45195-159, Iran\\
Max-Planck-Institut f\"ur Physik komplexer Systeme, N\"othnitzer Str.38,
01187 Dresden, Germany
}
\pacs{75.10.Jm, 75.10.Pq, 75.40.Cx}

\begin{abstract}
We have studied the zero temperature phase diagram of XYZ model
in the presence of transverse magnetic field.
We show that small anisotropy ($0\leq\Delta<1$) is not relevant to change
the universality class. The phase diagram
consists of two antiferromagnetic ordering
and a paramagnetic phases. We have obtained the critical exponents,
fixed points and running of coupling constants by implementing the
standard quantum renormalization group.
The continuous phase transition from antiferromagnetic (spin-flop)
phase to a paramagnetic one is in the universality class of
Ising model in transverse field.
Numerical exact diagonalization has been done to justify our results.
We have also addressed on the application of our findings
to the recent experiments on $Cs_2CoCl_4$.

\end{abstract}
\maketitle

Systems near criticality are usually characterised by fluctuations over
many length scales.
At the critical point itself, fluctuations exist over
all scales. At moderate temperatures
quantum fluctuations are usually suppressed compared with the
thermal ones. However if temperature is near zero, quantum
fluctuations especially in the low-lying states dominate thermal ones
and strongly influence the critical behavior of system.
Zero-temperature (quantum) phase transition may occur in the area of
spin systems by applying noncommuting magnetic field which introduces
quantum fluctuations. Such a situation has been studied in the three
dimensional Ising ferromagnet $LiHoF_4$ in a transverse
magnetic field ~\cite{bitko}. However due to its high dimensionality, the
system behaves in a mean-field-like manner. In this paper we are going
to consider the one-dimensional XYZ model in the presence of a transverse
field where quantum fluctuations of symmetry breaking field play
an essential role. Generally Renormalization Group (RG) is the
proper method to give us the universal behavior at long wave lengths
where other methods fail to work accurately.

The spin-$(s=)\frac{1}{2}$ Hamiltonian of this model on
a periodic chain of N-sites is
\begin{equation}
H=\sum_{i=1}^N [J_x \x{i}\x{i+1}+J_y \y{i}\y{i+1}
+\Delta \z{i}\z{i+1} -h \x{i}],
\label{ham}
\end{equation}
where $J_x>0$ and $J_y>0$ are exchange couplings in the XY easy plane,
$0 \leq \Delta < 1$ is the anisotropy in Z direction which is in $J_y$ units and $h$ is proportional
to the transverse field. $\sigma^{\alpha}; \alpha=x, y, z$ are Pauli matrices.

When $h=0$, the XXZ model ($J_x=J_y$) is known to be solvable and
critical (gapless) while
$-1 \leq \Delta \leq 1$~\cite{yang}. The Ising regime is $\Delta > 1$
and $\Delta \leq -1$ is the ferromagnetic case.
Magnetic field in the anisotropy direction commutes with
the Hamiltonian ($h=0$) and extends the gapless region (quasi long range
order) to a border where a transition to paramagnetic phase takes place.
The model is still integrable and can be explained by
a conformal field theroy with central charge $c=1$ (Ref.[\onlinecite{alcaraz}] and
references therein).

In the case of XXZ model a transverse field
breaks the $U(1)$ symmetry of the Hamiltonian
to a lower, Ising-like, which develops a gap.
The ground state then has long range anti-ferromagnetic order
($0\leq \Delta < 1$). However due to non-zero
projection of order parameter on field axis it is a spin-flop N\'{e}el state.
In fact at a special field ($h_N=2\sqrt{2J_x(J_x+\Delta)}$)
the ground state is known exactly to be of
classical N\'{e}el type ~\cite{muller,mori1}.
Phase diagram, scaling of gap and
some of the low excited states at $h_N$ has been studied
in Ref.[\onlinecite{dmitriev}].
The gap vanishes at a critical field $h_c$, where a
transition to paramagnetic phase occurs.
Classical approach to this model reveals the mean field results
~\cite{kurmann} which is exact as $s\rightarrow \infty$. However
the study of critical region needs quantum fluctuations to be taken
into account.
Exact diagonalization ~\cite{mori} and Density Matrix Renormalization
Group (DMRG) ~\cite{capraro} gives us the properties of stable phases.
A bosonization approch to this model in certain limits leads to a nontrivial fixed point and
a gapless line which separates two gapped phases \cite{dutta}, moreover the
connection to the axial next-nearest neighbor Ising model (ANNNI) has been addressed.
The applicability of mean-field approximation has been studied by comparing
with the DMRG results of magnetization and structure factor\cite{caux}.
Recently the effect of longitudinal magnetic field on both Ising model in 
Transverse Field (TF) \cite{ovchinnikov} and XXZ model in TF has been
discussed \cite{dima2}.
Here we are going to present the phase diagram of XYZ model,
Eq.~(\ref{ham}), by means of RG flow of coupling constants to show
explicitly its universality class.

Apart from theoretical point of view,
recent experiments on $Cs_2CoCl_4$ in the presence of transverse magnetic
field can be explained by XYZ model with $\Delta=0.25$ ~\cite{kenzelmann}.
Using Quantum Renormalization Group (QRG)
we will show explicitly that the anisotropy is not relevant and
the universality class is governed by Ising model in Transverse
Field (ITF). In addition  QRG results
rule out the existence of  spin liquid phase between spin flop
and paramagnetic phases which are separated
at the critical field ($h_c$). Exact diagonalization data supports
our QRG results by calculating the structure factor and magnetization
of finite chain sizes. Our results are in good agreement with the
experimental data. We will also discuss on the reasons why magnetization
does not saturate just above critical point.

Quantum RG scheme in real space is started by decomposing lattice into
isolated blocks. The Hamiltonian of each block is diagonalized exactly and
some of the low-lying states is kept to construct the basis for renormalized
Hilbert space. Finally the Hamiltonian is projected to the renormalized
space ~\cite{pfeuty}. We have considered a two sites block and kept
the ground ($\ket{\epsilon0}$) and first ($\ket{\epsilon1}$)
excited states of each block to construct the
embedding operator
($T=\ket{\epsilon1}\bra{\uparrow}+\ket{\epsilon0}\bra{\downarrow}$) ~\cite{miguel}.
Energy eigenvalues are $\epsilon0=-J_x-J_y-\Delta$ and
$\epsilon1=J_x-\sqrt{4h^2+(J_y-\Delta)^2}$. The $\ket{\uparrow}$ and
$\ket{\downarrow}$  are renamed basis in the renormalized Hilbert space.
The interaction between blocks define the effective interaction of
renormalized chain where each block is considered as a single site.
A remark is in order when projecting the Hamiltonian to the effective
(renormalized) Hilbert space. The effective Hamiltonian is not exactly
similar to the initial one, i.e. the sign of $\y{i}\y{i+1}$ and
$\z{i}\z{i+1}$ terms is changed. To avoid this and producing a self similar
Hamiltonian we first implement
a $\pi$ rotation around x-axis for
even sites and leaves odd sites unchanged.
Therefore the Hamiltonian is transformed to the following form,
\begin{equation}
H=\sum_{i=1}^{N/2} [J_x \x{i}\x{i+1}-J_y \y{i}\y{i+1}
-\Delta \z{i}\z{i+1} -h \x{i}].
\label{effham}
\end{equation}
We note to interpret our final
results in terms of this transformation.
The renormalized Hamiltonian ($H^{ren}=T^{\dagger} H(transformed) T$)
is similar to Eq.(\ref{effham}) with renormalized coupling defined below.
\begin{eqnarray}
J'_x&=&\frac{J_x}{4}\Big(\frac{(J_y-\Delta)^2-\vartheta^2}
{(J_y-\Delta)^2+\vartheta^2}\Big)^2
 \nonumber \\
J'_y&=&\frac{J_y}{2}\frac{(J_y-\Delta+\vartheta)^2}{(J_y-\Delta)^2+\vartheta^2}
\nonumber \\
h'&=&\frac{\epsilon0-\epsilon1}{2}-
\frac{J_x}{2}\Big(\frac{(J_y-\Delta)^2-\vartheta^2}
{(J_y-\Delta)^2+\vartheta^2}\Big)^2 \nonumber \\
\Delta'&=&\frac{\Delta}{2}\frac{(J_y-\Delta-\vartheta)^2}
{(J_y-\Delta)^2+\vartheta^2}
\label{rgflow}
\end{eqnarray}
where $\vartheta=\sqrt{4h^2+(J_y-\Delta)^2}-2h$.
This RG-flow is not valid when $h\rightarrow 0$
where the U(1) symmetry at $J_x=J_y$ can not be recovered
by Eq.~(\ref{rgflow}).
It will be discussed later. However due to level crossing
which happens for the eigenstates of block Hamiltonian,
Eq.~(\ref{rgflow}) is valid when $g_x\leq(1+\sqrt{1+4g_h^2})/2$ and
$g_{\Delta}\leq g_x \leq 1$. This covers XYZ model ($J_x \leq J_y$)
in transverse
field when $0\leq\Delta < 1$. The new parameters $g_x=\frac{J_x}{J_y}$,
$g_{\Delta}=\frac{\Delta}{J_y}$ and $g_h=\frac{h}{J_y}$ are defined because
these ratios actually define competing phases.

\begin{figure}
\centerline{\includegraphics[width=8cm,angle=0]{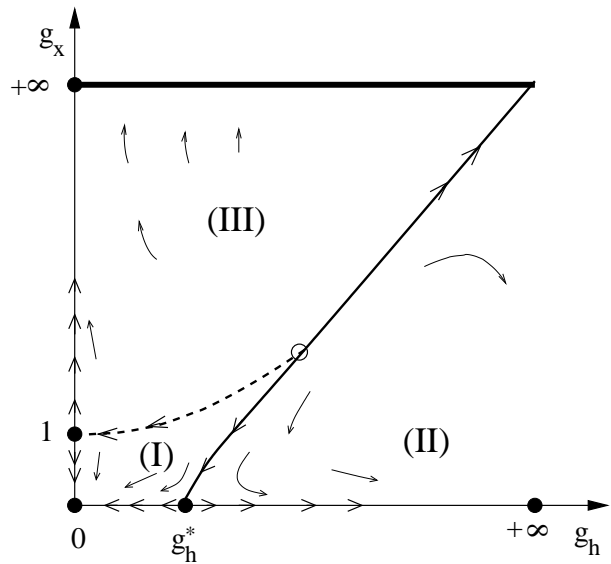}}
\caption{Phase diagram of XY model in transverse field. Arrows show
running of couplings under RG. Filled circles show fixed points and
the open circle is the tricritical point where
lines separating different phases merge. 
The tick line at the top of phase diagram ($g_x=+\infty$) is a line of fixed points \cite{langari2}.
Phase (I) is antiferromagnetic Ising in y-direction (spin-flop),
(II) paramagnetic in x-direction
and (III) is antiferromagnet in x-direction. $g_h^*$ is the ITF fixed point.}
\label{f.1}
\end{figure}

\begin{figure}
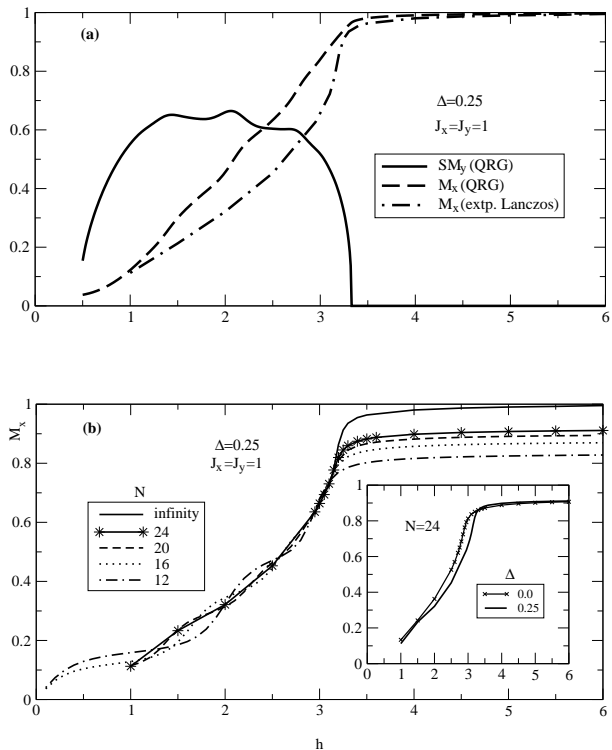

\centerline{\includegraphics[width=8cm,angle=0]{fig2a.eps}}
\vspace{5mm}
\centerline{\includegraphics[width=8cm,angle=0]{fig2b.eps}}
\caption{{\bf a} The order parameter ($SM_y$) and magnetization in
x-direction ($M_x$) versus transverse field. QRG and extrapolated
Lanczos results are compared for ($M_x$).
{\bf b} Lanczos results of $M_x$ vs. transverse field for
$N=12, 16, 20, 24$ and extrapolation to $N\rightarrow \infty$,
at $\Delta=0.25$ and $J_x=J_y=1$. The inset shows that $M_x$ behaves
qualitatively the same for $\Delta~=~0 ~\mbox{ and } 0.25$.}
\label{f.2}
\end{figure}

We have plotted the RG-flow (arrows) and different phases in Fig.~\ref{f.1}.
The RG equations (Eq.~(\ref{rgflow})) show  running of $\Delta$ to zero.
In other words the anisotropy term is irrelevant ($0\leq\Delta < 1$).
So we have only plotted the $\Delta=0$ plane. It means
that the universality class of XYZ model in transverse field (TF) is the
same as XY model in TF. Moreover the exchange interaction in the
x-direction is also irrelevant while $J_x<J_y$. As $J_x$ vanishes under
RG, there are only two effective terms in the Hamiltonian.
This is exactly the case of Ising model in TF (ITF model). So the interplay
of $J_y \y{i}\y{i+1}$ and $h (\x{i}+\x{i+1})$ defines either ordering
in y or paramagnetic in x direction. Solving the RG equation for fixed points,
we found the non-trivial fixed point
$g_h^*\equiv(g_x=0, g_h\simeq1.26, g_{\Delta}=0)$
apart from the other which is at$ (g_x=0, g_h=\infty, g_{\Delta}=0)$
and represents saturated ferromagnet.
We have linearised the RG-flow at $g_h^*$ and found one relevant direction
(whose eigenvalue is larger than one). The eigenvalues and corresponding
eigenvectors of linearised RG at $g_h^*$ in
($g_x, g_h, g_{\Delta})$ space are:
$\ket{\lambda_1=1.59}=(0, 1, 0);
\ket{\lambda_2=0.31}=(1, 1.64, 0);
\ket{\lambda_3=0.46}=(0, 0.62, 1)$.
The relevant direction ($\ket{\lambda_1}$) is the horizontal line
passing  through $g_h^*$ and
$\ket{\lambda_2}$ is the tick line ending at $g_h^*$.
The critical exponents at this fixed point are
$\beta=0.41, \nu=1.48$ and $z=0.55$.
The discrepancies of exponents from exact values ($\beta=0.125, \nu=1$ and $z=1$)~\cite{pfeuty-itf}
are the result of 2-sites blocking, however these are exactly equal
to the exponents of ITF chain which is calculated by QRG~\cite{miguel}.
As far as $g_x\leq1$,
the control parameter is $g_h$. When $g_h<g_h^c$ (phase (I)),
the staggered magnetization
in y-direction ($SM_y$) is nonzero which is the order parameter to represents
the phase transition at $g_h^c$ (the line which ends at $g_h^*$).
However magnetization
in x-direction ($M_x$) is also nonzero which causes to consider this phase as
a spin flop phase. This is an Ising like phase which has a nonzero gap.
This gap is going to be closed at $g_h^c$ where the transition to paramagnetic
phase takes place. At this point the quantum fluctuation of TF destroys
the antiferromagnetic (AF) ordering completely.
The paramagnetic phase (II) appears at $g_h>g_h^c$
where spins are aligned in the
field direction and will be saturated in high TF. Note that the proper order
parameter for this phase transition is {\it staggered
magnetization in y-direction}. So it is not necessary to gain the saturation
value for $M_x$ just after $g_h^c$.
This also happens in ITF model.
We have plotted both $SM_y$ and $M_x$
in Fig.~\ref{f.2}(a). The comparison with Lanczos results
show very good qualitative agreement. Although it is not expected that
QRG gives good quantitative results we got fairly well agreement with Lanczos
results.

To discuss the behavior close to $h=0$,
we need to take into account the $U(1)$ symmetry in the QRG scheme.
So we will consider the XY model at $h=0$ and the effect
of TF is taken into account by perturbation. In this case
the only relevant parameter is $g_x$. Implementing a 3 sites blocking,
the RG flow is: $g'_x=g_x^3$, which has 2 stable $g_x^*=0, \infty$ and
an unstable fixed point $g_x^*=1$. The stable fixed points define two
AF Ising phases ordered in y-direction ($g_x^*=0$) and
x-direction ($g_x^*=\infty$). The $g_x^*=1$ is the critical point where
a transition occurs between two stable phases. Now the transverse field
is considered perturbatively which gives the following RG flow for $g_h$.
\begin{equation}
g'_h=\Big(\frac{2g_x\sqrt{1+g_x^2}-g_x^2}{1+g_x^2}\Big) g_h
\hspace{5mm};\hspace{5mm} g_h \rightarrow 0
\label{rgh0}
\end{equation}
The perturbation approach is justified since $g_h \rightarrow 0$.
For any value of $g_x$, Eq.~(\ref{rgh0}) leads to $g'_h < g_h$, which
means the direction of flow is toward the $g_x$ axis. As a result of QRG
at $g_h=0$ we expect to have a phase transition at small $g_h$ by changing
$g_x$ close to $g_x\simeq1$. The boundary of this phase transition is
shown by dashed line in Fig.~\ref{f.1} which is the gapless line 
reported in Ref.[\onlinecite{dutta}].
This line represents the phase
transition between phases (I) and (III), AF Ising
in y- and x-direction, respectively. 
As $g_x\rightarrow \infty$
($J_y\rightarrow 0$) the model behaves as an AF Ising
in a longitudinal magnetic field.
In this limit a first order phase transition at
$\frac{h}{J_x}=1$ divides the AF $\frac{h}{J_x}<1$
from paramagnetic $\frac{h}{J_x}>1$ phases.
A line of fixed points comes out of a 3-sites block QRG ~\cite{langari2} 
for $ \frac{h}{J_x}<1$ which has been shown as a tick line at the top of phase diagram (Fig.~\ref{f.1}).
Thus a line
with slope $\frac{g_x}{g_h}=1$ (as $J_y\rightarrow 0$)
constructs the boundary of phase transition between
(II) and (III). This phase transition is in the universality class of
AF Ising in a magnetic field. To complete the structure of
phase diagram we propose a tri-critical point (open circle in Fig.~\ref{f.1})
which is the coexistence point of three phases.
Still we do not have an RG equation at this point.

\begin{figure}
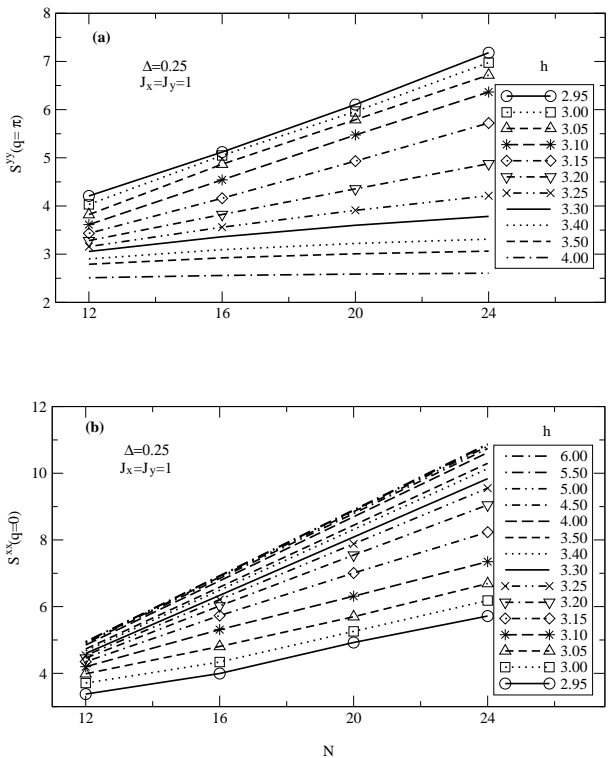

\centerline{\includegraphics[width=8cm,angle=0]{fig3a.eps}}
\vspace{6mm}
\centerline{\includegraphics[width=8cm,angle=0]{fig3b.eps}}
\caption{Structure factor {\bf (a)} $S^{yy}(q=\pi)$,
{\bf (b)} $S^{xx}(q=0)$
versus $N$ for different
transverse field. $S^{yy}(q=\pi)$ shows divergence as
$N\rightarrow\infty$ while $h<h_c\simeq3.1$ (in the ordered phase).
All plots for $S^{xx}(q=0)$ show divergence in thermodynamic limit
($N\rightarrow\infty$). However super linear behavior for $h<h_c\simeq3.1$
and almost linear behavior for $h>h_c$ is the sign of two different
phases.}
\label{f.3}
\end{figure}

We have implemented the Lanczos algorithm on finite sizes
($N=12, 16, 20, 24$) using periodic boundary conditions
to calculate $M_x$ and structure factors both in
x and y directions. In Fig.~\ref{f.2}(b) we have plotted $M_x$ for
different chain sizes at $\Delta=0.25$ and an extrapolation
to $N\rightarrow \infty$. The value of $\Delta=0.25$ is chosen to
fit the case of $Cs_2CoCl_4$. The general behavior is similar to what
we have obtained from QRG (Fig.~\ref{f.2}(a)). There is no sharp transition
to the saturation value at a given $h$ because $M_x$ is not the proper
order parameter to this phase transition.
Oscillations of $M_x$ at finite $N$ for $h<h_c$ is the result of
level crossing between ground and first excited states of this model.
The last level crossing happens at $h_N$.
We have also plotted the case of $\Delta=0$ to show the same qualitative
behavior as $\Delta=0.25$ in the inset of Fig.~\ref{f.2}(b).
Lanczos results leads to $SM_y=0$ for
any value of $h$, since in a finite system no symmetry breaking happens.
However the structure factor ($S^{yy}(q=\pi)$) diverges in the
ordered phase as $N\rightarrow\infty$. The structure factor at
momentum $q$ is defined as
\begin{equation}
S^{\alpha \alpha}(q)=\sum_{r} < \sigma_0^{\alpha} \sigma_r^{\alpha}>
e^{i q r}
\hspace{5mm};\hspace{5mm} \alpha=x, y
\label{sofq}
\end{equation}
In Fig.~\ref{f.3}(a), $S^{yy}(q=\pi)$ is plotted versus $N$ for different
transverse field. As far as $h>3.1$, $S^{yy}(q=\pi)$ grows slowly and
shows saturation at a finite value when $N\rightarrow\infty$.
In the other hand a super linear behavior versus $N$ shows a divergence
of structure factor for $h<3.1$. It corresponds to ordered phase which
is AF in y-direction. Thus the critical field at
$\Delta=0.25$ is $h_c=3.1\pm0.05$. A similar computation results to
$h_c=2.9\pm0.05$ for $\Delta=0$. To get an impression that
the QRG results are very surprising we just mention the value of
critical field for
comparison with Lanczos ones, $h_c(\Delta=0.25)=3.32$ and $h_c(\Delta=0)=3.12$.

We have also plotted the structure factor $S^{xx}(q=0)$ versus $N$
in Fig.~\ref{f.3}(b). This shows divergence for any value
of $h$ as $N\rightarrow\infty$ which verifies ordering in x-direction.
The spin flop phase (I) has nonzero $M_x$ which increases
by $h$ to the saturation value in paramagnetic phase (II). However
we observe different qualitative behavior for $h<h_c=3.1$ and
$h>h_c$. The former is super linear and the latter is  almost linear.
As mentioned before, $M_x$ is not the proper order parameter and is not
expected to be saturated at a specific $h$. The saturation happens for
enough large value of TF.

Summing up the QRG and numerical results, we claim that the universality
class of XYZ model in TF ($0\leq\Delta<1$) is the ITF model.
Thus there exists only two stable
phases, namely (I) and (II), which are distinguished by a critical field
at $h_c$. In this respect there is no spin liquid phase just after transition
point. We found very good agreement in the
sense of universal behavior with the experimental results ~\cite{kenzelmann} on $Cs_2CoCl_4$.
We have obtained the corresponding critical magnetic field $H_c=1.3^T$ comparing with
the reported $H_c=2.1^T$. The difference should come from two doublets
nature ($s=3/2$) of actual material and the effective Hamiltonian of $s=1/2$
in our calculation which is responsible for low fields.
The other mismatching is
the observed crossover behavior in $M_x$. As proposed
in Ref.~\cite{kenzelmann} the crossover behavior is related to
the saturation of the lower doublet of $Co^{2+}$ and the inset
of higher doublet effects. However for the XYZ chain as a spin 1/2 model
this does not happen. At $J_x=J_y$, applying small noncommuting fields
break the $U(1)$ rotational
symmetry and develops a gap which has the consequence of promoting
long-range order in a spin-flop phase (I). Increasing field stabilizes the
perpendicular AF order which can be observed by
the maximum in $SM_y$. Higher TF reduces ordering up to a critical
field ($h_c$) where gap vanishes. Just after this transition point
a gapped paramagnetic phase appears (II).


\acknowledgments
The author would like to thank  D. V. Dmitriev, V. Ya. Krivnov, 
A. A. Ovchinnikov, M. Peyravy, T. Vojta, K. Yang
and A. P. Young for fruitful discussions and useful comments.


\begin{thebibliography}{99}

\bibitem{bitko}
D. Bitko, T.F. Rosenbaum and G. Aeppli,
Phys. Rev. Lett. {\bf 77}, 940 (1996).


\bibitem{yang}
C.N. Yang and C.P. Yang,
Phys. Rev. {\bf 150}, 321 (1966); {\it ibid},{\bf 150}, 327, (1966).


\bibitem{alcaraz}
F. C. Alcaraz and A. L. Malvezzi,
J. Phys. A. {\bf 28}, 1521 (1995).

\bibitem{muller}
G. M\"uller and R. E. Shrock,
Phys. Rev. {\bf B32}, 5845 (1985).

\bibitem{mori1}
S. Mori, I. Mannari  and I. Harda,
J. Phys. Soc. Jpn. (63), 3474 (1994).

\bibitem{dmitriev}
D. V. Dmitriev, V. Ya. Krivnov, A. A. Ovchinnikov  and A. Langari,
{\bf JETP}, {\bf 95}, 538 (2002);
D. V. Dmitriev, V. Ya. Krivnov, and A. A. Ovchinnikov,
Phys. Rev. {\bf B 65}, 172409 (2002).


\bibitem{kurmann}
J. Kurmann, H. Thomas and G. M\"uller,
Physica {\bf A 112}, 235 (1982).


\bibitem{mori}
S. Mori, J.-J. Kim  and I. Harda,
J. Phys. Soc. Jpn. {\bf 64}, 3409 (1995).



\bibitem{capraro}
F. Capraro and C. Gros,
Eur. Phys. J. {\bf B 29}, 35 (2002).

\bibitem{dutta}
A. Dutta and D. Sen,
Phys. Rev. {\bf B 67}, 94435 (2003).

\bibitem{caux}
J.-S. Caux, F. H. L. Essler and U. L\"ow,
Phys. Rev. {\bf B 68}, 134431 (2003).

\bibitem{ovchinnikov}
A. A. Ovchinnikov, D. V. Dmitriev, V. Ya. Krivnov and O. O. Cheranovskii,
Phys. Rev. {\bf B 68}, 214406 (2003).

\bibitem{dima2}
D. V. Dmitriev and V. Ya. Krivnov, condmat/0403035.

\bibitem{kenzelmann}
M. Kenzelmann, R. Coldea, D. A. Tennant, D. Visser, M. Hofmann,
P. Smeibidl and Z. Tylczynski,
Phys. Rev. {\bf B 65}, 144432 (2002).

\bibitem{pfeuty}
P. Pfeuty P., R. Jullien and K. L. Penson,
{\it Real space renormalization, Chap. 5,
Ed. T. W. Burkhardt and J. M. J. van Leeuwen}
(Springer, Berlin, 1982).

\bibitem{miguel}
M. A. Martin-Delgado and G. Sierra,
Int. J. Mod. Phys. {\bf A 11}, 3145 (1996).

\bibitem{pfeuty-itf}
P. Pfeuty, Ann. Phys. {\bf 57}, 79 (1970).

\bibitem{langari2}
A. Langari, Phys. Rev. {\bf B 58}, 14467 (1998).




\end{thebibliography}
\end{document}